# Complementary Capabilities of Photoacoustic Imaging to Existing Optical Ocular Imaging Techniques

Dipen Kumar[a, *], Anju Goyal[b], Alan Truhan[c], Gary Abrams[b], Rayyan Manwar[d]

[a] Wayne State University School of Medicine, Detroit, MI, United States

[b] Department of Ophthalmology, Visual and Anatomical Sciences, Wayne State University School of Medicine, Detroit, MI, United States

[c] Wayne State University Physician Group, Kresge Eye Institute, Detroit, MI, United States

[d] Department of Biomedical Engineering, Wayne State University, Detroit, MI, United States

*Corresponding author:* dipen.kumar@med.wayne.edu

Since 1886, when the first picture of the human retina was taken, ocular imaging has played a crucial role in the diagnosis and management of ophthalmic diseases[1]. One of the biggest contributors to the advancement of ocular imaging is the adoption of optical imaging techniques. Optical imaging is a method of looking into the body in a non-invasive way, like x-rays. However, unlike radiological imaging techniques that use ionizing radiation, optical imaging uses light and the properties of photons to produce detailed images ranging from structures as small as cells and molecules to structures as large as tissues and organs. There are plenty of advantages of using optical imaging compared to radiological imaging. For one, optical imaging is much safer for patients since it uses non-ionizing radiation to excite electrons without causing damage. Additionally, since it is fast and safe, optical imaging can be used to monitor acute and chronic diseases, as well as treatment outcomes. Optical imaging is also useful for imaging soft tissue since different types of tissue absorb and scatter light differently. Finally, optical imaging can advantageously use varying colors of light to see and measure multiple properties of tissues at a time. Therefore, it is no surprise that the optical imaging modalities of fundus photography in the 1920's[2], scanning laser ophthalmoscope (SLO) imaging in 1981[3] and optical coherence tomography (OCT) in 1991[4] has touted a "golden age" in ophthalmic imaging applications[5]. Although these technologies have advanced the field of ocular imaging and are commonly used in clinical practice, they are not without their flaws. A new technology, photoacoustic imaging, has shown to have promising features that could make it the next major imaging technique in ophthalmology. Additionally, photoacoustic imaging (PA) can combine with preexisting optical microscopic imaging modalities to achieve multimodal imaging of the eye. In this chapter, we will give a brief overview of fundus photography, SLO, and OCT while discussing PA's potential as the next major ocular imaging modality.

First introduced in 1920 and extensively used since 1960, fundus photography continues to be a staple technique in ophthalmology[2]. Initially 35 mm film was once the standard for fundus photography but it has long been replaced by digital acquisition[5]. Fundus photography works in a similar fashion as an indirect ophthalmoscope. Light is focused by a series of lenses in a ringed-shaped aperture, which then is passed into a central aperture to form a ring which then passes through the camera objective lens and cornea to illuminate the retina. The reflected light from the retina then passes through a dark hole in the annulus formed by the illumination system previously described. There is minimal reflection of the light source in the captured image because the light rays of the two systems are independent. A picture can then be taken by using one mirror to interrupt the light from the illumination system so that the light from a flash bulb can pass into the eye. Another mirror drops at the same time, in front of the observation telescope to direct the reflected light onto film or a digital charged couple device (CCD). Monochromatic light can also be used rather than white light since monochromatic light allows for increased contrast of anatomical details of the fundus[6]. Normally fundus photography can only capture a small field of view (FOV) while the pupil is dilated, but it can be increased with a small aperture stop at the cost of resolution[2]. The maximum field of view is 50° but it can be increased to 60° if using a mydriatic camera[2]. Additionally, by using a special *Montage* software, individual images can be put together to form a collage that can cover up to 110°[2]. Furthermore, fundus photography can be combined with wide angle imaging to achieve a field of view between 45° and 140°, but there is proportionally less retinal magnification[5]. The main advantages of fundus photography are ease of use, full color, low cost compared to other imaging techniques and its high patient compliance[2]. Currently, fundus photography is used to monitor the progression of diseases like diabetic retinopathy, age-related macular degeneration (ARMD), glaucoma, and neoplasms of the eye[5].

In SLO, the retina is scanned in a rectangular pattern of parallel scanning lines followed by the electron beam on a TV or computer screen (raster pattern)[2] using a monochromatic, narrow laser beam. The beam is usually deflected using one slow vertical and one fast horizontal galvanometer scanner[7]. By modulating the scanning beam, projection of graphics in the raster is achieved. Since it uses a raster pattern, early SLOs could output to a TV monitor and be recorded on videotapes. The SLO has been further improved by combining other technology with it. Confocal scanning laser ophthalmoscope (cSLO) combines the principles of confocal imaging to increase contrast and depth resolution. Confocal microscopy was invented in 1955 by Marvin Minsky[8]. Confocal microscopy uses a pinhole (confocal filter), which is in an optically conjugate plane in front of a detector and point illumination to remove out-of-focus signal[2]. Much of the light that is reflected is blocked by the pinhole since light is only reflected by structures closer to the focal plane. 2D imaging occurs in a raster pattern over the specimen but 3D imaging is possible by changing the axial resolution. By increasing the numerical aperture or decreasing the diameter of the pinhole one can increase the depth. One can then scan many thin sections through a sample which can be combined with

SLO to allow cSLO to acquire depth information[9]. An improvement to cSLO is multi-spectral SLOs that use multiple lasers of different wavelengths. These lasers tend to be coaxial via a couple of dichroic combining mirrors and the goal is to introduce color to match images from fundus photography. The lasers either are multiplexed or fired simultaneously to an X-Y scanning mirror that causes the light to focus on a several millimeter-sized square on the retina. The reflected light then traverses to a beam splitter that directs a portion of the light to the detector[2]. Multispectral SLOs are used for retinal vessel oximetry, reflectometry, angioscotometry, and fundus perimetry[10-14]. Overall cSLO, is advantageous compared to previous imaging techniques since it allows for better images, patient comfort, video capability and the ability to image pupils that do not dilate well. It has been shown to be effective in detecting biomarkers for diabetic retinopathy[15], age-related macular degeneration[16], scanning the nerve head in glaucoma[17] and imaging the retinal nerve fiber layer (RNFL)[18]. The most common use of the SLO is with ultra-wide-field imaging of 200 degrees using the Optos System. This uses a SLO with an ellipsoidal lens to visualize the peripheral retina. About 82% of the retina can be imaged. Advantages include low light level for patient comfort and good images can often be obtained without dilation of the pupil. Fundus autofluorescence, fluorescein angiography and Indocyanine green angiography can be done with the Optos system. A more recent adaptation to SLO is adaptive optics SLO (AOSLO). Adaptive optics was a technology originally created for astronomy that has been combined with SLO to reduce the effects of wavefront distortions caused by optical aberrations. This is done by measuring the wavefront distortions and compensating for them by using devices such as a deformable mirror[19]. These distortions diminish the quality of the image being reflected by the eye which prevented microscopic resolution of structures such as capillaries and cells[3]. AOSLO most commonly uses a Shack-Hartmann sensor to measure these distortions by calculating the local phase errors in the wavefront. A phase modulator, such as a deformable mirror, can be used to correct these errors since the phase errors can be used to reconstruct the wavefront which in turn can control the deformable mirror. Another aspect to have a high magnification of small structures is image stabilization. Recently, eye tracking and stimulus delivery method have been implemented in AOSLO to achieve it[20].

Optical Coherence Tomography (OCT) is a non-invasive, micron level, high-resolution imaging technique based on the principal of Michelson interferometry that provides real-time images of the retina. As is with Michelson interferometry, an interference pattern is produced by splitting light into two arms: a sample arm from scanning the retina and a reference arm from a mirror. These arms are then recombined by semitransparent mirrors and redirected to a photodetector or camera[21]. If the interference is constructive between the two arms, the signal is strong at the detector and if they are destructive, the signal is weak at the detector. A reflectivity profile, also called an A-scan, can be gathered by scanning the mirror in the reference arm which contains information on the spatial dimensions and location of structures in the retina.

A cross-sectional tomograph, otherwise known as a B-scan, can be attained by combining a series of A-scans. OCT uses low coherence interferometry as opposed to conventional interferometry that uses long coherence length[22]. Low coherence interferometry uses low coherence light which is light that consists of a finite bandwidth of frequencies rather than just a single frequency. The broadband light allows for low coherence interferometry to shorten the interference to micrometers, perfect for its usage in ophthalmology. Additionally, it should be noted that OCT usually utilizes near-infrared (NIR) light since the relatively long wavelength allows for NIR to penetrate deeper than cSLO into scattering media like the retina. Since its inception in 1991, OCT has made huge advancements and improvements to increase the rate of imaging and resolution of OCT. Time Domain OCTs (TD OCT) have largely been replaced by Spectral Domain or Fourier-domain OCT (SD-OCT) since current state of the art ones can produce between 40 and 70,000 A-scans per minute, which is much faster than TD OCTs[5]. The major advantages of it being faster are that the scan takes less time and it is less impacted by artifacts and aberrations caused by blinking or eye movement[5]. Like SLO, OCT has been combined with adaptive optics (AO-OCT) to decrease the aberrations caused by imperfections in the curvature of the cornea and lens[23]. Plus, AO-OCT has the advantage of higher axial resolution compared to AO-SLO[23]. OCT used to be limited by the fact that it could not be used for blood flow analysis due to a poor delineation of blood vessels from the scattering of light as erythrocytes move through them[24]. However, three types of OCT have shown promise in this regard: Doppler OCT, OCT angiography (OCTA) and visible light OCT (vis-OCT). Doppler OCT combines OCT with the principles of the Doppler effect which results in improved resolution and sensitivity that allows for evaluation of blood flow, the volume of retinal and choroidal vasculature, and abnormalities in choroidal vasculature[25], and abnormalities in retinal and choroid vessels[26]. OCTA came about due to the improvements in OCT sensitivity and speed over the years which has led to better delineation of blood vessels[27]. OCTA compares consecutive B-scans taken at rates of several hundred Hz. The advantages to OCTA are that there is no need for the use of fluorescein dyes such as sodium fluorescein and indocyanine green [28], the ability for repeated scans, and the ability to analyze flow in a specific axial location of the retina or choroid[29]. Vis-OCT, which uses visible light rather than NIR has also recently gained attention since it has been shown to have better axial resolution than NIR based OCTs and it can have better image contrast due to tissues scattering properties in visible light, albeit at the cost of image depth[30]. On top of visualizing 3-D retinal structure, vis-OCT can quantify blood oxygen saturation ($sO_2$) in retinal circulation[25]. Due to its ability to show cross-sections of tissue layers at micrometer resolution, OCT is heavily used in ophthalmology as a method to assess structural changes in the retina in diseases such as diabetic retinopathy, vein occlusion, age-related macular degeneration, glaucoma, multiple sclerosis and other diseases that have ocular sequelae. OCT is very sensitive in detecting macular edema and is more accurate than clinical examination. OCT has significantly reduced false positive referrals for diabetic macular edema (DME) during diabetic screenings[31]. Additionally, OCT has given insight into abnormalities at the juncture between vitreous and the macula in patients with DME which could influence management and prognosis[32]. Furthermore, OCT is also useful in early uveitic macular edema detection[33] with the identification of specific OCT patterns associated with the disease[34]. Another disease that OCT is used for is age-related macular degeneration (ARMD). Fluorescein angiography has been largely replaced by OCT as the imaging method for monitoring ARMD treatment and the need for further anti-VEGF treatment[35]. OCT is also heavily used in cases of glaucoma. Glaucoma progression is associated with RNFL and ganglion cell thinning[36] so OCT can be used for glaucoma detection and progression[37]. While most of OCT technology is focused on imaging of the retina or pathologies related to the retina, enhanced depth imaging OCT (EDI-OCT) can evaluate choroidal thickness and posterior segment inflammatory disorders[5]. Aside from monitoring the choroid, it has been shown to be useful in monitoring other ocular inflammatory diseases such as Vogt Koyanagi Harada's disease[38], sarcoidosis[39], birdshot chorioretinopathy[40] and infectious choroiditis[41]. However, OCT has been well established in ophthalmology, it has also been used in other medical disciplines such as dermatology[42-55].

While fundus photography, SLO and OCT are still consistently used today in ophthalmology, they are not without their problems and limitations. To start, fundus photography requires pupil dilation with short-acting mydriatic drops which can cause discomfort for patients[5]. There have been recent advancements in cameras that do not require mydriatic drops but these can be affected by media opacity, such as cataracts, so mydriatic cameras are still the cameras of choice. Mydriatic cameras are especially desired if there is a need to image the periphery of the retina[56]. Even more so than discomfort to patients, these technologies suffer from a lack of quantitative data, lack of ability to take photographs of high quality, poor depth resolution, difficulty in comparing serial photographs and the need to subject patient's to high-intensity light to illuminate the retina[2]. As for SLO, one of the limitations is that involuntary eye movements affect image quality. A solution to this is Tracking SLO (TSLO) which uses a high-speed retinal tracker to significantly improve image quality[56]. Another limitation to SLO is current commercial SLOs, such as Optos or the Heidelberg wide lens, do not provide images of the eye from ora to ora[57]. Additionally, there is a distortion of the image on the periphery of the image since it is taking a 2D image of a 3D globe[58]. Also, the measurements of the eye, such as distance and area, may not be the actual dimensions of the eye since it does not standardize the image to any axis of the eye[5]. Artifacts on the image can also be caused by several things: eyelashes, cataracts, intraocular lens implants, pigments in the anterior segment of the eye and vitreous opacities to name a few[59]. Furthermore, the cost of equipment and maintenance of SLO can be a large barrier[5]. Finally, there are the limitations to OCT. OCT by itself is unable to measure $sO_2$ and RPE melanin. While OCTA exists, has a restricted by its limited field of view, lack of information on fill or flow speed, and motion artifacts[60]. Vis-OCT suffers from limited image depth and it can cause discomfort for eye imaging[23]. Finally, since all three techniques are optical scattering-based modalities, measurements of blood oxygen saturation in the eye are affected by light scattering, and fundus photography and SLO also need to use contrast agents to measure them[61].

When light is received by the eye, it is processed by both the retinal pigment epithelium (RPE) and the retina which consumes a large amount of oxygen and energy[62]. Therefore, the retina needs supporting vasculature which it has from retinal and choroidal circulation. Normally these vasculature systems bring oxygen and nutrients to the retina[63], and studies have shown that variations in the $sO_2$ and RPE melanin play a role in ocular diseases such as diabetic retinopathy[64], glaucoma[65], retinal venous occlusion[66] and ARMD[67]. Thus, there has been an increased effort in the past decade to quantify the $sO_2$ and RPE melanin concentration in the eye. Fortunately, both blood and melanin, within the visible light spectral range, have high optical absorption coefficients which allow them to be measured[68]. Photoacoustic imaging (PAI) has been shown to measure optical absorption properties of both blood and melanin in a non-invasive and precise way in other locations of the body[69,70]. Therefore, PAI is a recent technology for ophthalmology due to its potential clinical use in measuring retinal and choroidal $sO_2$ and the RPE melanin. Photoacoustic imaging has been well studied in several preclinical imaging applications[71-80]. It is based on the photoacoustic effect, which is the generation of ultrasound waves due to the absorption of light and thermal expansion[81]. The primary PA imaging technique is photoacoustic tomography (PAT). PAT starts by using a laser to illuminate and excite the sample where, short(nanosecond) laser pulses are used that satisfy the stress and thermal confinements. The sample then exhibits the photoacoustic effect as it absorbs energy from the laser which results in heat emission, transient thermoelastic expansion and leads to ultrasound wave generation[69,70]. The generated acoustic wave is detected by ultrasound transducers and recorded as a function of time which then is converted based on the sound speed in the sample into a one-dimensional depth-resolved image, also called an A-line. By aligning the A-lines based on their spatial location, a transverse linear scan of the point laser illumination on the sample can make a 2D image. From there a 2D raster scan of the point of illumination creates a 3D image. PAT can be categorized into photoacoustic computed tomography (PACT) or photoacoustic microscopy (PAM). PACT uses an array of ultrasonic transducers (multiple single element, linear, phased, ring, circular, or spherical arrays) to detect PA waves emitted from an object at multiple view angles[82] while PAM uses the raster scanning method[83]. Even though a higher penetration depth can be achieved using PACT, it comes at the expense of coerce resolution, system and computational costs[84]. On the other hand, higher resolution PAM systems can be classified based on their spatial resolution or the type of scanning they use with limited penetration depth. For spatial resolution, PAM systems can either be acoustic-resolution where the imaging resolution is based on the focus of the ultrasonic detector[85] or optical resolution where the resolution is determined by the optical focal spot[86]. As for the scanning classifications, there is mechanical-scanning which simultaneously translates the optical illumination and ultrasound detection for volumetric imaging[87] and optical-scanning where there is a set of galvanometers which maintain the ultrasound detection stationary while they scan a focused optical illumination[87]. Currently, PA is capable of imaging structures in both the anterior and posterior segments of the eye. Originally, it had been used to examine ocular structures such as the iris or retinal vasculature qualitatively[88], but current PA imaging now focus on quantification of properties like $sO_2$[87] or retinal oxygen metabolic rate ($rMRO_2$)[89] in the eye. The major structure that PA imaging currently focuses on in the anterior segment of the eye is the iris, specifically the red blood cells in the microvasculature and melanin of the iris[62]. While both mechanical-scanning acoustic (AR-PAM) and optical resolution PA microscopy (OR-PAM) have been used to image the iris, only mechanical-scanning OR-PAM has been able to obtain high-resolution images of iris microvasculature[90]. The system works by focusing laser illumination light onto the iris microvasculature using a microscope objective lens[91]. A water tank is placed over the subject's eye so that a focused ultrasonic detector can receive the ultrasonic signals emitted from the iris[91]. Additionally, $sO_2$ of the iris microvasculature can be measured by using two excitation wavelengths that have different oxy-hemoglobin and deoxy-hemoglobin absorption coefficients[91]. Iris melanin has also been measured by PA imaging using mechanical-scanning OR-PAM[91]. However, unlike the $sO_2$ of the iris microvasculature only qualitative measuring of iris melanin has been performed[84]. Instead of the iris, the focus of PA imaging in the posterior segment is the red blood cells in the retinal and choroidal microvasculature along with melanin in the RPE[62]. Both mechanical-scanning OR-PAM and AR-PAM have been used to image the posterior segment of the eye[92], but the resolution is too low to visualize the microvasculature in AR-PAM[92] and in OR-PAM the lens attenuates the ultrasonic signals resulting in reduced SNR of the images[93]. To overcome this, optical-scanning PA microscopy (OSPAM) was developed[87]. Unlike mechanical-scanning, OSPAM uses a pulsed laser coupled to a 1 x 2 single-mode optical fiber[87]. One of the outputs allowed for compensation of laser intensity variation, while the other was directed to the cornea using a pair of galvanometer mirrors and a pair of telescope lenses[87]. Additionally, OSPAM uses an ultrasonic needle transducer to detect PA waves, thus eliminating the need for a water tank[62]. Moreover, the needle prevents major signal attenuation resulting in high SNR images[62]. Lastly, although it has not been used in PA imaging of the eye, contrast agents improve PA image quality[94] and extend the scope of PA imaging to the genetic and molecular level[95]. Some of the contrast agents used like Evans blue[96], indocyanine green, Indocyanine-green-embedded PEBBLEs as a contrast agent for photoacoustic imaging

[97] and nanoparticles[98] are already common ophthalmic contrast agents thus inviting the possibility of using them with ocular PAM.

Unfortunately, while PA imaging shows a lot of promise as an upcoming ocular imaging modality, it is relatively new and has many limitations that need to be addressed before the clinical translation. Firstly, photoacoustic signals detection requires physical contact with the eye. Whether it's a water tank or a needle transducer with ultrasonic gel, both cause patient discomfort and are not suitable for clinical settings[62]. Additionally, physical motion for saccades or head movement can disrupt PA imaging. While there have been strides taken to fix this problem, there are many concerns about the performance stability and detection sensitivity with these non-contact PA methods[62]. Secondly, OSPAM still requires extended imaging depth for both the retina and the choroid, high resolutions for RPE melanin and fast imaging speeds to reduce motion artifacts. For depth, optical clearing agents could be used[99] but they are not usable for *in vivo* imaging and NIR light could be used but the high-power excitation is a safety concern[100]. For improving the resolution of PA, one could potentially increase the lateral resolution by using the synthetic aperture technique[101]. As for axial resolution, a broad ultrasonic bandwidth does increase the axial resolution however higher sensitivity in OSPAM is achieved with narrower bandwidth[102]. A balance needs to be determined to maximize both axial resolution and detection sensitivity. Finally, higher imaging speed could reduce motion artifacts and while increasing the laser repetition rate can increase imaging speed, it is limited by the ultrasound propagation time from the posterior eye. Lastly, before PA imaging can be clinically adapted, it requires numerous animal studies to confirm the longitudinal performance stability of PA measurements in the eye. Furthermore, there is limited knowledge of PA imaging for early ocular disease detection[62]. Finally, studies have shown that visual stimulation of the retina can result in changes to retinal vessel diameter, blood flow and $sO_2$[103]. Therefore, studies need to be done on how visual light illumination affects OSPAM accuracy.

The biggest advantage of OSPAM is that multimodal imaging is achievable by combining OSPAM with other imaging modalities. The development of multimodal microscopic imaging techniques has become increasingly important in the biomedical community since it allows for comprehensive physiological information of biological tissues[104]. In the case of ocular imaging, most optical image modalities work by detecting the scattering of light reflected from the eye or fluorescent light stimulated in the sample. The problem is that these modalities require the back-traveling of photons from the sample, so they cannot measure the optical absorption. Therefore, OSPAM complements these modalities well because it is currently the only optical absorption-based imaging modality[84]. Thus, by combining the two, one can get anatomical information, like cellular layer organization of the retina, from preexisting ocular imaging techniques and molecular information, like $sO_2$, from OSPAM which gives a quantitative, holistic image of the eye. OSPAM can be combined with auto-florescence imaging[105], fluorescein angiography[65], SLO[84] and most importantly OCT. OCT adds to OSPAM by allowing for detailed, high resolution, retinal and choroidal structural information[105]. Additionally, by using repeated OCT scanning, complete retinal vasculature mapping is possible[106]. Furthermore, OCT can quantitatively measure retinal blood flow rate and velocity by detecting the Doppler phase shifts produced by moving blood[107]. Finally, OCT can be used to guide OSPAM so that an area of interest on posterior segment can be imaged[108].

Ocular imaging has come a long way since the first image of the retina in 1886. The addition of optical imaging modalities to ophthalmology has introduced faster and more precise methods for physicians to monitor and diagnose ocular pathologies. While fundus photography, SLO, and OCT have advanced ocular imaging to a large degree, they have clear limitations in being optical scattering-based imaging modalities shown in Table 1. Therefore, the introduction of photoacoustic imaging to ophthalmology could lead to the development of a novel, stand-alone modality and/or a complimentary modality to OCT and SLO that could advance the field of ocular imaging.

Table 1. List of Ophthalmological Imaging Modalities and Their Applications, Advantages and Limitations

| Technology | Applications | Advantages | Limitations |
| --- | --- | --- | --- |
| **Fundus Photography** | Retinal fundus imaging, diabetes, ARMD, glaucoma, neoplasms of the eye | Quick and simple technique to master, true view of the retina, observes a larger retinal field at any one time compared with ophthalmoscopy, high patient compliance, able to monitor progression of diseases and low cost compared to other imaging modalities | Image produced is 2D, difficulty observing and assessing abnormalities due to lack of depth appreciation on images, less magnification and image clarity, conditions such as cataracts reduce image clarity, artifact errors may produce unusual images |
| **SLO** | Retinal vessel oximetry, reflectometry, | High lateral resolution, fast imaging, high quality images, patient | Low depth resolution, high maintenance cost, affected by motion |

| | angioscotometry, fundus perimetry diabetic retinopathy, age-related macular degeneration, scanning the nerve head in glaucoma and imaging the retinal nerve fiber layer | comfort and video capability | artifacts, distortion of image at the periphery and light scattering affects sO2 |
|---|---|---|---|
| **OCT** | Macular edema, macular degeneration, glaucoma, multiple sclerosis | High lateral and depth resolution | Based OCT has poor delineation of blood vessels and limited field of view in OCT angiography |
| **PAOM** | sO2 and RPE imaging | Optical absorption based, medium depth perception and multimodal imaging with other modalities | Only optical absorption imaging, currently requires physical contact, needs more testing before clinically available |

**References**

1. Taruttis, A.; Ntziachristos, V. Advances in real-time multispectral optoacoustic imaging and its applications. *Nature Photonics* **2015**, *9*, 219.
2. Gramatikov, B.I. Modern technologies for retinal scanning and imaging: an introduction for the biomedical engineer. *Biomedical engineering online* **2014**, *13*, 52.
3. Webb, R.H.; Hughes, G.W. Scanning laser ophthalmoscope. *IEEE Transactions on Biomedical Engineering* **1981**, 488-492.
4. Huang, D.; Swanson, E.A.; Lin, C.P.; Schuman, J.S.; Stinson, W.G.; Chang, W.; Hee, M.R.; Flotte, T.; Gregory, K.; Puliafito, C.A. Optical coherence tomography. *science* **1991**, *254*, 1178-1181.
5. Bajwa, A.; Aman, R.; Reddy, A.K. A comprehensive review of diagnostic imaging technologies to evaluate the retina and the optic disk. *International ophthalmology* **2015**, *35*, 733-755.
6. Lin, D.Y.; Blumenkranz, M.S.; Brothers, R.J.; Grosvenor, D.M.; Group, T.D.D.S. The sensitivity and specificity of single-field nonmydriatic monochromatic digital fundus photography with remote image interpretation for diabetic retinopathy screening: a comparison with ophthalmoscopy and standardized mydriatic color photography. *American journal of ophthalmology* **2002**, *134*, 204-213.

7. Webb, R.H. Optics for laser rasters. *Applied optics* **1984**, *23*, 3680-3683.
8. Minsky, M. Memoir on inventing the confocal scanning microscope. *Scanning* **1988**, *10*, 128-138.
9. Vieira, P.; Manivannan, A.; Lim, C.; Sharp, P.; Forrester, J. Tomographic reconstruction of the retina using a confocal scanning laser ophthalmoscope. *Physiological measurement* **1999**, *20*, 1.
10. Vieira, P.; Manivannan, A.; Sharp, P.F.; Forrester, J.V. True colour imaging of the fundus using a scanning laser ophthalmoscope. *Physiological measurement* **2001**, *23*, 1.
11. Elsner, A.E.; Burns, S.A.; Hughes, G.W.; Webb, R.H. Reflectometry with a scanning laser ophthalmoscope. *Applied optics* **1992**, *31*, 3697-3710.
12. Remky, A.; Beausencourt, E.; Elsner, A.E. Angioscotometry with the scanning laser ophthalmoscope. Comparison of the effect of different wavelengths. *Investigative ophthalmology & visual science* **1996**, *37*, 2350-2355.
13. Lompado, A.; Smith, M.H.; Hillman, L.W.; Denninghoff, K.R. Multispectral confocal scanning laser ophthalmoscope for retinal vessel oximetry. In Proceedings of Spectral imaging: Instrumentation, applications, and analysis; pp. 67-74.
14. Remky, A.; Elsner, A.E.; Morandi, A.J.; Beausencourt, E.; Trempe, C.L. Blue-on-yellow perimetry with a scanning laser ophthalmoscope: small alterations in the central macula with aging. *JOSA A* **2001**, *18*, 1425-1436.
15. Wykes, W.; Pyott, A.; Ferguson, Y. Detection of diabetic retinopathy by scanning laser ophthalmoscopy. *Eye* **1994**, *8*, 437.
16. Manivannan, A.; Kirkpatrick, J.; Sharp, P.; Forrester, J. Clinical investigation of an infrared digital scanning laser ophthalmoscope. *British journal of ophthalmology* **1994**, *78*, 84-90.
17. Seymenoğlu, G.; Başer, E.; Öztürk, B. Comparison of spectral-domain optical coherence tomography and Heidelberg retina tomograph III optic nerve head parameters in glaucoma. *Ophthalmologica* **2013**, *229*, 101-105.
18. Chan, E.W.e.; Liao, J.; Foo, R.C.M.; Loon, S.C.; Aung, T.; Wong, T.Y.; Cheng, C.-Y. Diagnostic performance of the ISNT rule for glaucoma based on the Heidelberg retinal tomograph. *Translational vision science & technology* **2013**, *2*, 2-2.
19. Liang, J.; Williams, D.R.; Miller, D.T. Supernormal vision and high-resolution retinal imaging through adaptive optics. *JOSA A* **1997**, *14*, 2884-2892.
20. Burns, S.A.; Tumbar, R.; Elsner, A.E.; Ferguson, D.; Hammer, D.X. Large-field-of-view, modular, stabilized, adaptive-optics-based scanning laser ophthalmoscope. *JOSA A* **2007**, *24*, 1313-1326.
21. Puliafito, C.A.; Hee, M.R.; Lin, C.P.; Reichel, E.; Schuman, J.S.; Duker, J.S.; Izatt, J.A.; Swanson, E.A.; Fujimoto, J.G. Imaging of macular diseases with optical coherence tomography. *Ophthalmology* **1995**, *102*, 217-229.
22. Fercher, A.; Mengedoht, K.; Werner, W. Eye-length measurement by interferometry with partially coherent light. *Optics letters* **1988**, *13*, 186-188.
23. Pircher, M.; Zawadzki, R.J. Review of adaptive optics OCT (AO-OCT): principles and applications for retinal imaging. *Biomedical optics express* **2017**, *8*, 2536-2562.
24. Drexler, W.; Liu, M.; Kumar, A.; Kamali, T.; Unterhuber, A.; Leitgeb, R.A. Optical coherence tomography today: speed, contrast, and multimodality. *Journal of biomedical optics* **2014**, *19*, 071412.
25. Izatt, J.A.; Kulkarni, M.D.; Yazdanfar, S.; Barton, J.K.; Welch, A.J. In vivo bidirectional color Doppler flow imaging of picoliter blood volumes using optical coherence tomography. *Optics letters* **1997**, *22*, 1439-1441.
26. Leitgeb, R.A.; Schmetterer, L.; Hitzenberger, C.K.; Fercher, A.F.; Berisha, F.; Wojtkowski, M.; Bajraszewski, T. Real-time measurement of in vitro flow by Fourier-domain color Doppler optical coherence tomography. *Optics letters* **2004**, *29*, 171-173.

27. Ang, M.; Tan, A.C.; Cheung, C.M.G.; Keane, P.A.; Dolz-Marco, R.; Sng, C.C.; Schmetterer, L. Optical coherence tomography angiography: a review of current and future clinical applications. *Graefe's Archive for Clinical and Experimental Ophthalmology* **2018**, *256*, 237-245.
28. Gao, S.S.; Jia, Y.; Zhang, M.; Su, J.P.; Liu, G.; Hwang, T.S.; Bailey, S.T.; Huang, D. Optical Coherence Tomography Angiography. *Investigative ophthalmology & visual science* **2016**, *57*, OCT27-OCT36, doi:10.1167/iovs.15-19043.
29. Keane, P.A.; Sadda, S.R. Retinal imaging in the twenty-first century: state of the art and future directions. *Ophthalmology* **2014**, *121*, 2489-2500.
30. Shu, X.; Beckmann, L.J.; Zhang, H.F. Visible-light optical coherence tomography: a review. *Journal of biomedical optics* **2017**, *22*, 121707.
31. Koizumi, H.; Pozzoni, M.C.; Spaide, R.F. Fundus autofluorescence in birdshot chorioretinopathy. *Ophthalmology* **2008**, *115*, e15-e20.
32. Otani, T.; Kishi, S.; Maruyama, Y. Patterns of diabetic macular edema with optical coherence tomography. *American journal of ophthalmology* **1999**, *127*, 688-693.
33. Hassenstein, A.; Bialasiewicz, A.A.; Richard, G. Optical coherence tomography in uveitis patients. *American journal of ophthalmology* **2000**, *130*, 669-670.
34. Markomichelakis, N.N.; Halkiadakis, I.; Pantelia, E.; Peponis, V.; Patelis, A.; Theodossiadis, P.; Theodossiadis, G. Patterns of macular edema in patients with uveitis: qualitative and quantitative assessment using optical coherence tomography. *Ophthalmology* **2004**, *111*, 946-953.
35. Krebs, I.; Ansari-Shahrezaei, S.; Goll, A.; Binder, S. Activity of neovascular lesions treated with bevacizumab: comparison between optical coherence tomography and fluorescein angiography. *Graefe's Archive for Clinical and Experimental Ophthalmology* **2008**, *246*, 811-815.
36. Bussel, I.I.; Wollstein, G.; Schuman, J.S. OCT for glaucoma diagnosis, screening and detection of glaucoma progression. *British Journal of Ophthalmology* **2014**, *98*, ii15-ii19.
37. Jeoung, J.W.; Choi, Y.J.; Park, K.H.; Kim, D.M. Macular ganglion cell imaging study: glaucoma diagnostic accuracy of spectral-domain optical coherence tomography. *Investigative ophthalmology & visual science* **2013**, *54*, 4422-4429.
38. Maruko, I.; Iida, T.; Sugano, Y.; Oyamada, H.; Sekiryu, T.; Fujiwara, T.; Spaide, R.F. Subfoveal choroidal thickness after treatment of Vogt–Koyanagi–Harada disease. *Retina* **2011**, *31*, 510-517.
39. Modi, Y.S.; Epstein, A.; Bhaleeya, S.; Harbour, J.W.; Albini, T. Multimodal imaging of sarcoid choroidal granulomas. *Journal of ophthalmic inflammation and infection* **2013**, *3*, 58.
40. Keane, P.A.; Allie, M.; Turner, S.J.; Southworth, H.S.; Sadda, S.R.; Murray, P.I.; Denniston, A.K. Characterization of birdshot chorioretinopathy using extramacular enhanced depth optical coherence tomography. *JAMA ophthalmology* **2013**, *131*, 341-350.
41. Goldenberg, D.; Goldstein, M.; Loewenstein, A.; Habot-Wilner, Z. Vitreal, retinal, and choroidal findings in active and scarred toxoplasmosis lesions: a prospective study by spectral-domain optical coherence tomography. *Graefe's Archive for Clinical and Experimental Ophthalmology* **2013**, *251*, 2037-2045.
42. Hojjatoleslami, A.; Avanaki, M.R.N. OCT skin image enhancement through attenuation compensation. *Applied Optics* **2012**, *51*, 4927-4935, doi:10.1364/AO.51.004927.
43. Avanaki, M.R.N.; Cernat, R.; Tadrous, P.J.; Tatla, T.; Podoleanu, A.G.; Hojjatoleslami, S.A. Spatial Compounding Algorithm for Speckle Reduction of Dynamic Focus OCT Images. *IEEE Photonics Technology Letters* **2013**, *25*, 1439-1442, doi:10.1109/LPT.2013.2266660.
44. Hojjatoleslami, S.; Avanaki, M.; Podoleanu, A.G. Image quality improvement in optical coherence tomography using Lucy–Richardson deconvolution algorithm. *Applied optics* **2013**, *52*, 5663-5670.

45. Avanaki, M.R.; Podoleanu, A.G.; Schofield, J.B.; Jones, C.; Sira, M.; Liu, Y.; Hojjat, A. Quantitative evaluation of scattering in optical coherence tomography skin images using the extended Huygens–Fresnel theorem. *Applied optics* **2013**, *52*, 1574-1580.
46. Avanaki, M.R.N.; Hojjatoleslami, A.; Sira, M.; Schofield, J.B.; Jones, C.; Podoleanu, A.G. Investigation of basal cell carcinoma using dynamic focus optical coherence tomography. *Applied Optics* **2013**, *52*, 2116-2124, doi:10.1364/AO.52.002116.
47. Hojjat, A.; Podoleanu, A.G. Investigation of computer-based skin cancer detection using optical coherence tomography AU - Avanaki, Mohammad R.N. *Journal of Modern Optics* **2009**, *56*, 1536-1544, doi:10.1080/09500340902990007.
48. Adabi, S.; Hosseinzadeh, M.; Noie, S.; Daveluy, S.; Clayton, A.; Mehregan, D.; Conforto, S.; Nasiriavanaki, M. Universal in vivo Textural Model for Human Skin based on Optical Coherence Tomograms. *arXiv preprint arXiv:1706.02758* **2017**.
49. Avanaki, M.R.N.; Hojjatoleslami, A. Skin layer detection of optical coherence tomography images. *Optik* **2013**, *124*, 5665-5668, doi:https://doi.org/10.1016/j.ijleo.2013.04.033.
50. Adabi, S.; Turani, Z.; Fatemizadeh, E.; Clayton, A.; Nasiriavanaki, M. Optical coherence tomography technology and quality improvement methods for optical coherence tomography images of skin: a short review. *Biomedical engineering and computational biology* **2017**, *8*, 1179597217713475.
51. Taghavikhalilbad, A.; Adabi, S.; Clayton, A.; Soltanizadeh, H.; Mehregan, D.; Avanaki, M.R.N. Semi-automated localization of dermal epidermal junction in optical coherence tomography images of skin. *Applied Optics* **2017**, *56*, 3116-3121, doi:10.1364/AO.56.003116.
52. Faiza, M.; Adabi, S.; Daoud, B.; Avanaki, M.R.N. High-resolution wavelet-fractal compressed optical coherence tomography images. *Applied Optics* **2017**, *56*, 1119-1123, doi:10.1364/AO.56.001119.
53. Avanaki, M.R.; Podoleanu, A. En-face time-domain optical coherence tomography with dynamic focus for high-resolution imaging. *Journal of biomedical optics* **2017**, *22*, 056009.
54. Turani, Z.; Fatemizadeh, E.; Blumetti, T.; Daveluy, S.; Moraes, A.F.; Chen, W.; Mehregan, D.; Andersen, P.E.; Nasiriavanaki, M. Optical Radiomic Signatures Derived from Optical Coherence Tomography Images to Improve Identification of Melanoma. *Cancer Research* **2019**, 10.1158/0008-5472.Can-18-2791, canres.2791.2018, doi:10.1158/0008-5472.Can-18-2791.
55. Adabi, S.; Fotouhi, A.; Xu, Q.; Daveluy, S.; Mehregan, D.; Podoleanu, A.; Nasiriavanaki, M. An overview of methods to mitigate artifacts in optical coherence tomography imaging of the skin. *Skin Research and Technology* **2018**, *24*, 265-273.
56. Bennett, T.J.; Barry, C.J. Ophthalmic imaging today: an ophthalmic photographer's viewpoint–a review. *Clinical & experimental ophthalmology* **2009**, *37*, 2-13.
57. Hammer, D.X.; Ferguson, R.D.; Magill, J.C.; White, M.A.; Elsner, A.E.; Webb, R.H. Compact scanning laser ophthalmoscope with high-speed retinal tracker. *Applied optics* **2003**, *42*, 4621-4632.
58. Chou, B. Limitations of the panoramic 200 Optomap. *Optometry and vision science* **2003**, *80*, 671-672.
59. Spaide, R.F. Peripheral areas of nonperfusion in treated central retinal vein occlusion as imaged by wide-field fluorescein angiography. *Retina* **2011**, *31*, 829-837.
60. Dunphy, R.W.; Wentzolf, J.N.; Subramanian, M.; Conlin, P.R.; Pasquale, L.R. Structural features anterior to the retina represented in panoramic scanning laser fundus images. *Ophthalmic Surgery, Lasers and Imaging Retina* **2008**, *39*, 160-163.
61. Zhang, M.; Hwang, T.S.; Campbell, J.P.; Bailey, S.T.; Wilson, D.J.; Huang, D.; Jia, Y. Projection-resolved optical coherence tomographic angiography. *Biomedical optics express* **2016**, *7*, 816-828.

62. Liu, W.; Zhang, H.F. Photoacoustic imaging of the eye: a mini review. *Photoacoustics* **2016**, *4*, 112-123.
63. Yu, D.-Y.; Cringle, S.J. Oxygen distribution and consumption within the retina in vascularised and avascular retinas and in animal models of retinal disease. *Progress in retinal and eye research* **2001**, *20*, 175-208.
64. Campochiaro, P.A. Molecular pathogenesis of retinal and choroidal vascular diseases. *Progress in retinal and eye research* **2015**, *49*, 67-81.
65. Hardarson, S.H.; Stefánsson, E. Retinal oxygen saturation is altered in diabetic retinopathy. *British journal of ophthalmology* **2012**, *96*, 560-563.
66. Olafsdottir, O.B.; Hardarson, S.H.; Gottfredsdottir, M.S.; Harris, A.; Stefánsson, E. Retinal oximetry in primary open-angle glaucoma. *Investigative ophthalmology & visual science* **2011**, *52*, 6409-6413.
67. Hardarson, S.H.; Stefansson, E. Oxygen saturation in central retinal vein occlusion. *American journal of ophthalmology* **2010**, *150*, 871-875.
68. Berendschot, T.T.; Goldbohm, R.A.; Klopping, W.A.; van de Kraats, J.; van Norel, J.; van Norren, D. Influence of lutein supplementation on macular pigment, assessed with two objective techniques. *Investigative ophthalmology & visual science* **2000**, *41*, 3322-3326.
69. Jacques, S.L. Optical properties of biological tissues: a review. *Physics in Medicine & Biology* **2013**, *58*, R37.
70. Xu, M.; Wang, L.V. Photoacoustic imaging in biomedicine. *Review of scientific instruments* **2006**, *77*, 041101.
71. Nasiriavanaki, M.; Xia, J.; Wan, H.; Bauer, A.Q.; Culver, J.P.; Wang, L.V. High-resolution photoacoustic tomography of resting-state functional connectivity in the mouse brain. *Proceedings of the National Academy of Sciences* **2014**, *111*, 21-26, doi:10.1073/pnas.1311868111.
72. Yao, J.; Xia, J.; Maslov, K.I.; Nasiriavanaki, M.; Tsytsarev, V.; Demchenko, A.V.; Wang, L.V. Noninvasive photoacoustic computed tomography of mouse brain metabolism in vivo. *Neuroimage* **2013**, *64*, 257-266.
73. Mozaffarzadeh, M.; Mahloojifar, A.; Orooji, M.; Kratkiewicz, K.; Adabi, S.; Nasiriavanaki, M. Linear-array photoacoustic imaging using minimum variance-based delay multiply and sum adaptive beamforming algorithm. *Journal of biomedical optics* **2018**, *23*, 026002.
74. Xia, J.; Li, G.; Wang, L.; Nasiriavanaki, M.; Maslov, K.; Engelbach, J.A.; Garbow, J.R.; Wang, L.V. Wide-field two-dimensional multifocal optical-resolution photoacoustic-computed microscopy. *Optics Letters* **2013**, *38*, 5236-5239, doi:10.1364/OL.38.005236.
75. Mohammadi-Nejad, A.-R.; Mahmoudzadeh, M.; Hassanpour, M.S.; Wallois, F.; Muzik, O.; Papadelis, C.; Hansen, A.; Soltanian-Zadeh, H.; Gelovani, J.; Nasiriavanaki, M. Neonatal brain resting-state functional connectivity imaging modalities. *Photoacoustics* **2018**.
76. Mahmoodkalayeh, S.; Jooya, H.Z.; Hariri, A.; Zhou, Y.; Xu, Q.; Ansari, M.A.; Avanaki, M.R.N. Low Temperature-Mediated Enhancement of Photoacoustic Imaging Depth. *Scientific Reports* **2018**, *8*, 4873, doi:10.1038/s41598-018-22898-2.
77. Meimani, N.; Abani, N.; Gelovani, J.; Avanaki, M.R. A numerical analysis of a semi-dry coupling configuration in photoacoustic computed tomography for infant brain imaging. *Photoacoustics* **2017**, *7*, 27-35.
78. Rayyan Manwar, M.H., Ali Hariri, Karl Kratkiewicz, Shahryar Noei, Mohammad R. N. Avanaki Photoacoustic Signal Enhancement: Towards Utilization of Low Energy Laser Diodes in Real-Time Photoacoustic Imaging. *Photoacoustic Sensing and Imaging in Biomedicine, MDPI Sensors* **2018**.

79. Mohammadi, L.; Behnam, H.; Nasiriavanaki, M. Modeling skull's acoustic attenuation and dispersion on photoacoustic signal. In Proceedings of Photons Plus Ultrasound: Imaging and Sensing 2017; p. 100643U.
80. Zafar, M.; Kratkiewicz, K.; Manwar, R.; Avanaki, M. Development of Low-Cost Fast Photoacoustic Computed Tomography: System Characterization and Phantom Study. *Applied Sciences* **2019**, *9*, 374.
81. Yao, J.; Maslov, K.I.; Zhang, Y.; Xia, Y.; Wang, L.V. Label-free oxygen-metabolic photoacoustic microscopy in vivo. *Journal of biomedical optics* **2011**, *16*, 076003.
82. Fatima, A.; Kratkiewicz, K.; Manwar, R.; Zafar, M.; Zhang, R.; Huang, B.; Dadashzadesh, N.; Xia, J.; Avanaki, M. Review of Cost Reduction Methods in Photoacoustic Computed Tomography. *Photoacoustics* **2019**, 100137.
83. Hu, S.; Maslov, K.I.; Tsytsarev, V.; Wang, L.V. Functional transcranial brain imaging by optical-resolution photoacoustic microscopy. *Journal of biomedical optics* **2009**, *14*, 040503.
84. Cox, B.T.; Laufer, J.G.; Beard, P.C.; Arridge, S.R. Quantitative spectroscopic photoacoustic imaging: a review. *Journal of biomedical optics* **2012**, *17*, 061202.
85. Wang, L.V.; Gao, L. Photoacoustic microscopy and computed tomography: from bench to bedside. *Annual review of biomedical engineering* **2014**, *16*, 155-185.
86. Xu, M.; Wang, L.V. Universal back-projection algorithm for photoacoustic computed tomography. *Physical Review E* **2005**, *71*, 016706.
87. Xing, W.; Wang, L.; Maslov, K.; Wang, L.V. Integrated optical-and acoustic-resolution photoacoustic microscopy based on an optical fiber bundle. *Optics letters* **2013**, *38*, 52-54.
88. Maslov, K.; Zhang, H.F.; Hu, S.; Wang, L.V. Optical-resolution photoacoustic microscopy for in vivo imaging of single capillaries. *Optics letters* **2008**, *33*, 929-931.
89. Song, W.; Wei, Q.; Liu, T.; Kuai, D.; Zhang, H.F.; Burke, J.M.; Jiao, S. Integrating photoacoustic ophthalmoscopy with scanning laser ophthalmoscopy, optical coherence tomography, and fluorescein angiography for a multimodal retinal imaging platform. *Journal of biomedical optics* **2012**, *17*, 061206.
90. Jiao, S.; Jiang, M.; Hu, J.; Fawzi, A.; Zhou, Q.; Shung, K.K.; Puliafito, C.A.; Zhang, H.F. Photoacoustic ophthalmoscopy for in vivo retinal imaging. *Optics express* **2010**, *18*, 3967-3972.
91. Hennen, S.N.; Xing, W.; Shui, Y.-B.; Zhou, Y.; Kalishman, J.; Andrews-Kaminsky, L.B.; Kass, M.A.; Beebe, D.C.; Maslov, K.I.; Wang, L.V. Photoacoustic tomography imaging and estimation of oxygen saturation of hemoglobin in ocular tissue of rabbits. *Experimental eye research* **2015**, *138*, 153-158.
92. Hu, S.; Rao, B.; Maslov, K.; Wang, L.V. Label-free photoacoustic ophthalmic angiography. *Optics letters* **2010**, *35*, 1-3.
93. Thijssen, J.M.; Mol, H.J.M.; Timmer, M.R. Acoustic parameters of ocular tissues. *Ultrasound in Medicine and Biology* **1985**, *11*, 157-161, doi:10.1016/0301-5629(85)90018-3.
94. Luke, G.P.; Yeager, D.; Emelianov, S.Y. Biomedical Applications of Photoacoustic Imaging with Exogenous Contrast Agents. *Annals of Biomedical Engineering* **2012**, *40*, 422-437, doi:10.1007/s10439-011-0449-4.
95. Li, W.; Chen, X. Gold nanoparticles for photoacoustic imaging. *Nanomedicine (Lond)* **2015**, *10*, 299-320, doi:10.2217/nnm.14.169.
96. Yao, J.; Maslov, K.; Hu, S.; Wang, L.V. Evans blue dye-enhanced capillary-resolution photoacoustic microscopy in vivo. *Journal of biomedical optics* **2009**, *14*, 054049-054049, doi:10.1117/1.3251044.
97. Kim, G.; Huang, S.-W.; Day, K.C.; O'Donnell, M.; Agayan, R.R.; Day, M.A.; Kopelman, R.; Ashkenazi, S. Indocyanine-green-embedded PEBBLEs as a contrast agent for photoacoustic imaging. *Journal of Biomedical Optics* **2007**, *12*, 1-8, 8.

98. Yang, X.; Stein, E.W.; Ashkenazi, S.; Wang, L.V. Nanoparticles for photoacoustic imaging. *Wiley Interdisciplinary Reviews: Nanomedicine and Nanobiotechnology* **2009**, *1*, 360-368, doi:10.1002/wnan.42.
99. Zhou, Y.; Yao, J.; Wang, L.V. Optical clearing-aided photoacoustic microscopy with enhanced resolution and imaging depth. *Optics letters* **2013**, *38*, 2592-2595, doi:10.1364/OL.38.002592.
100. Hai, P.; Yao, J.; Maslov, K.I.; Zhou, Y.; Wang, L.V. Near-infrared optical-resolution photoacoustic microscopy. *Optics Letters* **2014**, *39*, 5192-5195, doi:10.1364/OL.39.005192.
101. Li, M.-L.; Zhang, H.F.; Maslov, K.; Stoica, G.; Wang, L.V. Improved in vivo photoacoustic microscopy based on a virtual-detector concept. *Optics Letters* **2006**, *31*, 474-476, doi:10.1364/OL.31.000474.
102. Ma, T.; Zhang, X.; Chiu, C.T.; Chen, R.; Kirk Shung, K.; Zhou, Q.; Jiao, S. Systematic study of high-frequency ultrasonic transducer design for laser-scanning photoacoustic ophthalmoscopy. *Journal of biomedical optics* **2014**, *19*, 16015-16015, doi:10.1117/1.JBO.19.1.016015.
103. Song, W.; Wei, Q.; Feng, L.; Sarthy, V.; Jiao, S.; Liu, X.; Zhang, H.F. Multimodal photoacoustic ophthalmoscopy in mouse. *Journal of biophotonics* **2013**, *6*, 505-512, doi:10.1002/jbio.201200061.
104. Jiao, S.Z., Hao F. . Multimodal microscopy for comprehensive tissue characterizations. *Advanced Biophotonics: Tissue Optical Sectioning, CRC Press* **2016**, 475-505.
105. Fernández, E.J.; Hermann, B.; Považay, B.; Unterhuber, A.; Sattmann, H.; Hofer, B.; Ahnelt, P.K.; Drexler, W. Ultrahigh resolution optical coherence tomography and pancorrection for cellular imaging of the living human retina. *Optics Express* **2008**, *16*, 11083-11094, doi:10.1364/OE.16.011083.
106. Jia, Y.; Tan, O.; Tokayer, J.; Potsaid, B.; Wang, Y.; Liu, J.J.; Kraus, M.F.; Subhash, H.; Fujimoto, J.G.; Hornegger, J., et al. Split-spectrum amplitude-decorrelation angiography with optical coherence tomography. *Optics express* **2012**, *20*, 4710-4725, doi:10.1364/OE.20.004710.
107. Liu, W.; Yi, J.; Chen, S.; Jiao, S.; Zhang, H.F. Measuring retinal blood flow in rats using Doppler optical coherence tomography without knowing eyeball axial length. *Medical Physics* **2015**, *42*, 5356-5362, doi:10.1118/1.4928597.
108. Zhang, H.F.; Puliafito, C.A.; Jiao, S. Photoacoustic ophthalmoscopy for in vivo retinal imaging: current status and prospects. *Ophthalmic Surg Lasers Imaging* **2011**, *42 Suppl*, S106-S115, doi:10.3928/15428877-20110627-10.